\documentclass[aip,jap,graphicx,reprint]{revtex4-1}

\usepackage{graphicx}

\begin{document}

\title{Overcoming device unreliability with continuous learning in a population coding based computing system} 

\author{Alice Mizrahi}
\email[]{alice.mizrahi@nist.gov}
\affiliation{National Institute of Standards and Technology, Gaithersburg, USA}
\affiliation{Maryland NanoCenter, University of Maryland, College Park, USA}
\author{Julie Grollier}
\affiliation{Unit\'e Mixte de Physique CNRS, Thales, Univ. Paris-Sud, Universit\'e Paris-Saclay, 91767, Palaiseau, France}
\author{Damien Querlioz}
\affiliation{Centre de Nanosciences et de Nanotechnologies, Univ. Paris-Sud, CNRS, Universit\'e Paris-Saclay, 91405, Orsay, France}
\author{M.D. Stiles}
\affiliation{National Institute of Standards and Technology, Gaithersburg, USA}

\date{\today}

\begin{abstract}
The brain, which uses redundancy and continuous learning to overcome the unreliability of its components, provides a promising path to building computing systems that are robust to the unreliability of their constituent nanodevices. In this work, we illustrate this path by a computing system based on population coding with magnetic tunnel junctions that implement both neurons and synaptic weights. We show that equipping such a system with continuous learning enables it to recover from the loss of neurons and makes it possible to use unreliable synaptic weights (\textit{i.e.} low energy barrier magnetic memories). There is a tradeoff between power consumption and precision because low energy barrier memories consume less energy than high barrier ones. For a given precision, there is an optimal number of neurons and an optimal energy barrier for the weights that leads to minimum power consumption.

\end{abstract}

\maketitle 

\section{Introduction}

The small size and low energy consumption of nanoelectronics devices make them ideal components for novel forms of computing. However, these advantages come with drawbacks: device to device variability, possibly stochastic behavior, and sometimes, device failure. Biological systems, like our brains, give hope that these drawbacks can be overcome because they operate in noisy environments with components, neurons and synapses, that are variable and normally exhibit stochastic behavior.\cite{faisal_noise_2008} Our brain is capable of recovery after massive loss of neurons, as can happen in an accident.\cite{mason_neurocognitive_2014} Two factors contribute to these abilities. First, the brain exhibits high redundancy, which is studied by neuroscientists within the framework of population coding. \cite{georgopoulos_primate_1988,pouget_information_2000,pouget_probabilistic_2013} Second, the brain never stops learning.\cite{merzenich_adaptable_1999,pascual-leone_plastic_2005,hofer_lifelong_2006} In contrast to most artificial neural networks in which the synaptic weights are set after an initial training phase,\cite{lecun_deep_2015} synaptic weights in the brain constantly adjust to adapt to changes. 

We show that equipping an artificial neural network with redundancy and, most importantly, continuous learning allows it to overcome the unreliability of nanodevices. We illustrate this idea with a system that uses population coding with stochastic magnetic tunnel junctions as spiking neurons and stable magnetic tunnel junctions to store synaptic weights.\cite{mizrahi_neural-like_2018} 
We first present the system and summarize the relevant physics of magnetic tunnel junctions. Numerical simulations of this system show that continuous learning enables it to recover from the catastrophic loss of neurons (caused by the breakdown of the tunnel barrier of junctions for instance), whether the loss occurs before or after training. With continuous learning, this type of system can learn in spite of unreliable synaptic weights. Here, magnetic tunnel junctions with energy barriers as low as $15 k_{\rm{B}}T$ lead to the same system precision as perfectly reliable weights. Because writing low barrier junctions consumes less energy than writing high barrier ones, there is a tradeoff between power consumption and precision. For a given precision, there is an optimal number of neurons and weights and an optimal barrier for the weight junctions, for which power consumption of the system is minimal.

\section{Description of the system}
\label{sec_coding}

\begin{figure}
\centering
\includegraphics[width=3.5in]{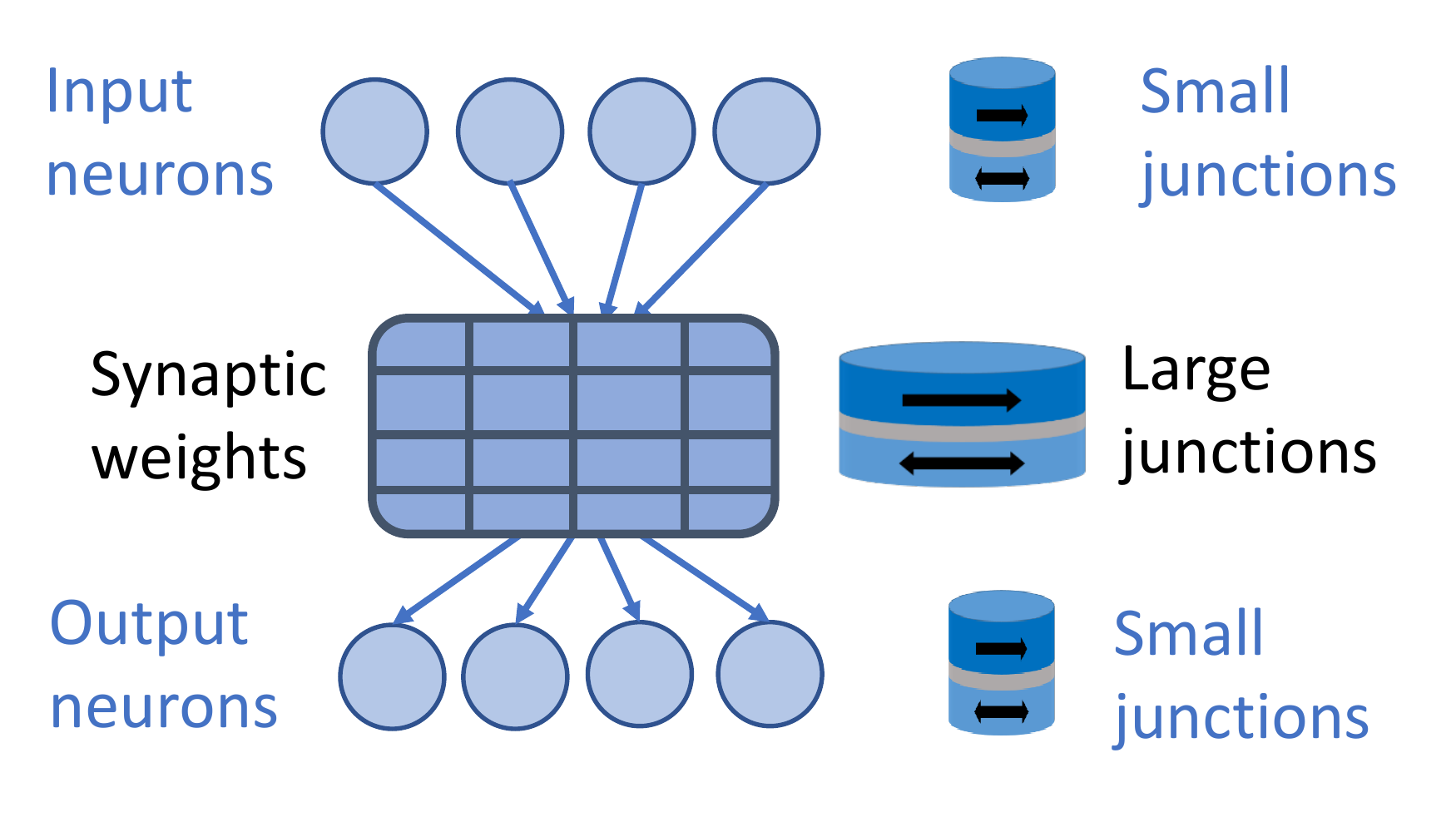}
\caption{Schematic of the considered system. Two neuron populations -- emulated by small unstable magnetic tunnel junctions -- are connected by an array of synaptic weights -- stored in large stable magnetic tunnel junctions.}
\label{fig_system}
\end{figure}

To illustrate the power of population coding, which provides intrinsic redundancy, we consider the computing system depicted in Fig. \ref{fig_system}.\cite{mizrahi_neural-like_2018} It is based on controllable-rate stochastic devices as neurons and (mostly) non-volatile devices to store synaptic weights as well as control circuitry realized with conventional electronic circuits. An analog input value is encoded by the firing rates of the population of spiking neurons.\cite{georgopoulos_primate_1988,pouget_information_2000,pouget_probabilistic_2013} This system can perform nonlinear transformations of the input value by connecting two populations linearly by the synaptic weights.\cite{salinas_transfer_1995} The rates of the output junctions, which encode the transformed value, satisfy:
\begin{equation}
<R^\mathrm{out}_j>~=~\sum_{i=1}^{N_\mathrm{in}}W_{ij}<R^\mathrm{in}_i>
\label{eq:rates}
\end{equation}
where $N_\mathrm{in}$ is the number of input junctions, $R^\mathrm{out}_j$ is the rate of the $j$-th output junction, $R^\mathrm{in}_i$ is the rate of the $i$-th input junction and $W_{ij}$ is the synaptic weight connecting the $i$-th input junction to the $j$-th output junction. The value encoded by the output population is a nonlinear function of the value encoded by the input population, and this function depends on the synaptic weights. Details about the encoding are given in the Appendix.


Due to their unlimited endurance, magnetic tunnel junctions are promising devices to implement such systems where learning is key. They consist of two thin ferromagnetic layers separated by a tunnel barrier. The state of the device can be read through its tunneling magneto resistance,\cite{yuasa_giant_2004,parkin_giant_2004} \textit{i.e.} each magnetic configuration (whether the magnetizations in the layers are parallel (P) or anti-parallel (AP)) has a different resistance. The probability to switch from one state to another after a given time $\Delta t$ is driven by a thermally activated Poisson process with escape rates $\phi_{\rm{P/AP}}$:\cite{rippard_thermal_2011,mizrahi_magnetic_2015}
\begin{equation}
P_{\rm{P/AP}}\left(\Delta t\right) = 1 - \exp\left( - \phi_{\rm{P/AP}} ~\Delta t \right)
\label{eq:proba}
\end{equation}
\begin{equation}
\phi_{\rm{P/AP}}=\phi_0 \exp\left(-\frac{\Delta E}{k_{\rm B}T}\left(1\pm \frac{V}{V_c}\right)\right)
\label{eq:model_smtj}
\end{equation}
where $\phi_0=10^9~\mathrm{s}^{-1}$ is the attempt frequency, $\Delta E$ is the energy barrier between the two stable states, $k_{\rm B}$ is the Boltzmann constant, $T$ is the temperature, $V_{\rm{c}}$ is the critical voltage and $V$ is the voltage applied across the junction. High energy barriers lead to long retention of the state: junctions with high energy barriers function as binary non-volatile memory, such as in Magnetic Random Access Memories.\cite{apalkov_magnetoresistive_2016} Low energy barriers cause the junction to randomly switch between the two states.\cite{rippard_thermal_2011,vodenicarevic_low-energy_2017} The switching rate can be controlled by applying a voltage across the junction through the phenomenon of spin transfer torque\cite{slonczewski_currents_2005,ralph_spin_2008}, as captured by the voltage dependent factor in Eq. \ref{eq:model_smtj}. This control of the spiking rate is reminiscent of sensory neurons and a key feature for population coding.\cite{mizrahi_neural-like_2018} 

Several recent approaches to novel computing use magnetic tunnel junctions in their stochastic regime.\cite{mizrahi_controlling_2016,sutton_intrinsic_2017,liyanagedera_stochastic_2017,camsari_stochastic_2017,fukushima_spin_2014,choi_magnetic_2014,parks_superparamagnetic_2017,lee_design_2017,rangarajan_spin-based_2017}
The present system uses one single stack of materials to implement both neurons and synapses, as depicted in Figure \ref{fig_system}. Small junctions have low barriers and emulate neurons, while large junctions have high barriers and store the synaptic weights (several junctions per weight are required since they are binary). 

\begin{figure}
\centering
\includegraphics[width=3in]{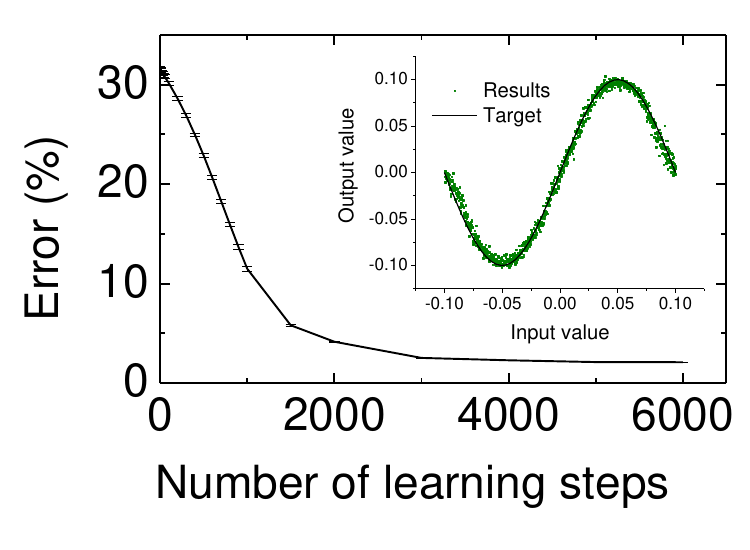}
\caption{Main panel: Error versus the number of learning steps for the \textit{sine} transformation. Each population is composed of 100 neurons. The error is computed as the distance between the output value and target value and expressed in percentage of the range of output values. Each data point corresponds to an average over 50 learning processes and for each of them the error was calculated as an average over 50 different input values. The error bar indicates the single standard deviation of the error in the mean, computed over the 50 learning processes. Inset: output value versus input value. The black solid line is the target \textit{sine} function and the green squares are the simulation results for a single learning process after 4050 learning steps.}
\label{fig_learning}
\end{figure}

Figure \ref{fig_learning} shows the results of numerical simulations illustrating learning with this system, for the case of perfectly reliable components. In this paper, we use a \textit{sine} transformation to demonstrate this system because it is non-linear and non-monotonic. Learning is done by trial and error. At first the weights are random, which gives an error of about $29\%$. At each learning step a random input value is chosen. If the output is close enough to the desired value, the weights are left unchanged, otherwise, they are modified according to a learning rule detailed in the Appendix. The learning is evaluated by computing the error, set to be the mean distance between the value encoded by the output population and its target divided by the range of possible output values. In Fig. \ref{fig_learning}, the error is plotted versus the number of learning steps. It progressively decreases down to a final minimum error of approximately $2\%$ of the output range. This minimum error depends among others on the number of junctions used and the range of voltages needed to tune the switching rates.\cite{mizrahi_neural-like_2018} Parameters used in these simulations are given in the Appendix.

Ref.~\onlinecite{mizrahi_neural-like_2018} uses circuit level simulations to demonstrate that a hardware implementation of this system would consume little energy only require a small area. Most importantly, the learning of the weights would consume sufficiently low energy and require sufficiently low area to be implemented on-chip and proceed continuously during the use of the system. The present results illustrate how continuous learning can overcome unreliability of components. 



\section{Robustness to the loss of neurons}

\begin{figure}
\centering
\includegraphics[width=3.5in]{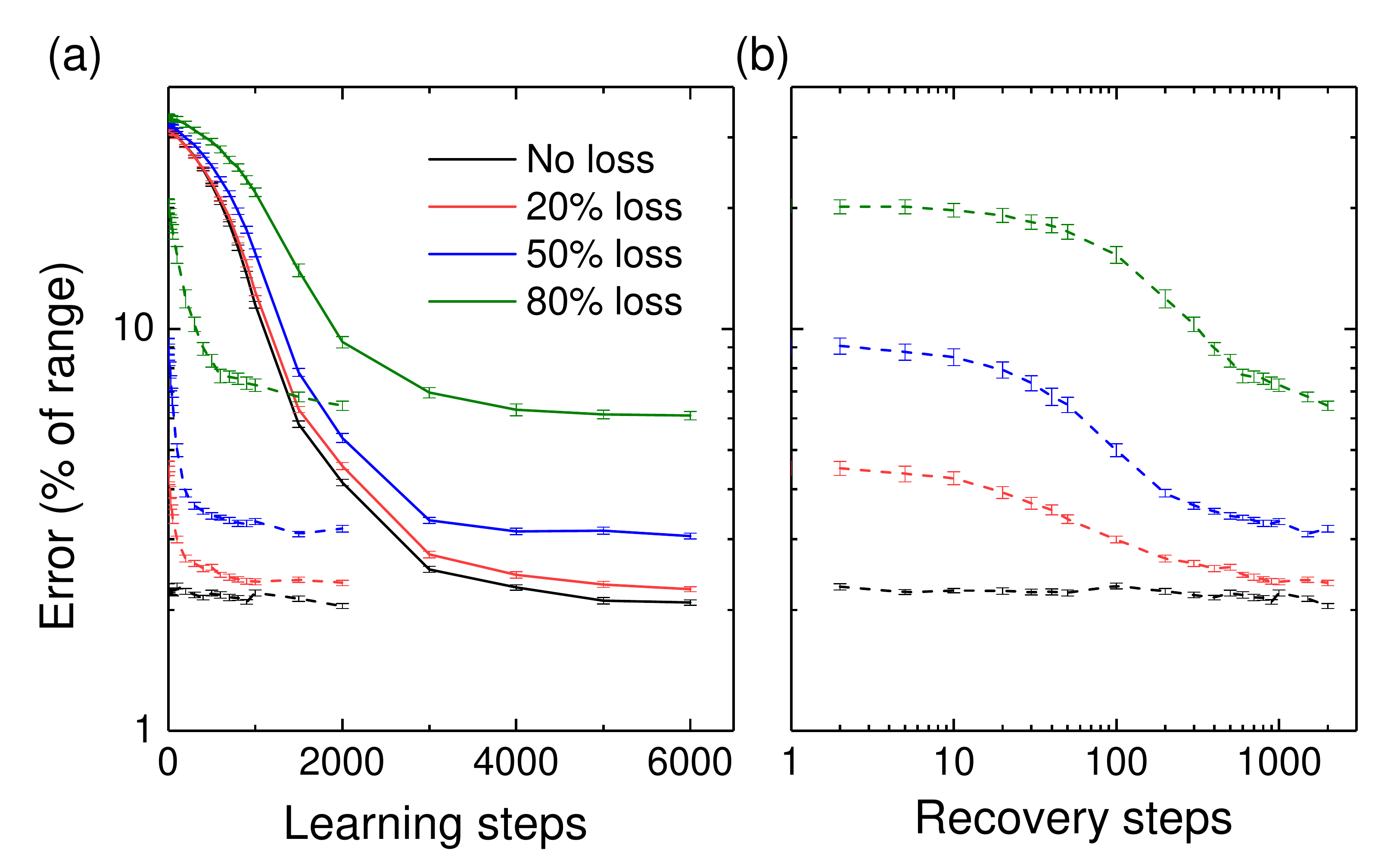}
\caption{Relative output error versus the number of learning steps. (a) Dashed lines: recovery after the loss of neurons. Solid lines: learning starting from a reduced population. (b) Recovery after the loss of neurons (enlargement of the dashed lines on panel (a)). The different colors correspond to different amounts of loss, on an initial system where input and output population each have 100 neurons. Each data point corresponds to an average over 50 learning instances and for each of them the error was estimated as an average over 50 different input values. The error bar indicates the single standard deviation of the error in the mean, computed over the 50 learning instances.}
\label{fig_neuron_loss}
\end{figure}

Magnetic tunnel junctions can occasionally fail, due to the breakdown of the tunnel barrier. 
To investigate the effects of such failure, we consider a system where both the input and output populations are composed of $N~=~$100 junctions. We model catastrophic component failure by selecting at random a fixed number of junctions in both the input and output populations and assigning them a null rate.
The solid lines in Fig. \ref{fig_neuron_loss}(a) show how the error evolves with the number of learning steps when neuron failure occurs before any learning. The different colors correspond to different loss levels and thus to different resulting population sizes. As expected from standard population coding works\cite{salinas_transfer_1995, abbott_decoding_1994} and Ref.~\onlinecite{mizrahi_neural-like_2018}, the final error is smaller for larger populations. 
The dashed lines in Fig. \ref{fig_neuron_loss}(a) and (b) show how the error evolves when neural loss occurs after initial learning. The system is initially trained with 4000 learning steps (reaching the steady state), and then, the neural loss is simulated. We observe that the loss increases the error of the system. However, this error is much lower than it was before the initial training (even with 80 \% loss, the error is less than half of what it is without training). 

After neuron loss, we continue to train the resulting system. We observe recovery, that is, the error decreases to a steady-state value that depends on the loss level. This steady-state value is the same as it would be for a system starting with that number of neurons as seen by comparing the recovery with the training of a system composed of a comparably reduced population of neurons, plotted as solid lines in Fig. \ref{fig_neuron_loss}(a). Furthermore, this state is reached much faster than through initial learning (a few hundred steps versus thousands). This rapid re-learning shows the ability of the system to adapt to radical changes by re-learning, building on previous knowledge. The results in this paper demonstrate how continuous learning makes our system resilient to the loss of neurons. 

\section{Robustness to unreliable synaptic weights}

The main reliability concern for the synaptic weights $W_{i,j}$, is the ability of the devices to retain information. In the case of magnetic tunnel junctions, reducing energy consumption requires lowering energy barriers, which reduces the long term non-volatility of these devices.
Quantifying the effect of this volatility  on performance is crucial to understanding how the system behaves. 
First, we study the initial learning process for different energy barrier heights in the devices storing the weights. Each finite barrier height $\Delta E_{\mathrm{w}}$ leads to a probability $P_{\mathrm{loss}}$ for the weight to lose its information at each learning step.
Because several binary junctions are necessary to store one weight, this loss probability is computed as follows:
\begin{equation}
P\left(\Delta t, \Delta E_{\mathrm{w}} \right) = 1 - \exp\left( - N_{\rm{bits}}~\Delta t~\phi_0 \exp\left( - \frac{\Delta E_{\mathrm{w}}}{k_{\mathrm{B}}T}  \right)\right)
\end{equation}
Where  $N_{\rm{bits}}$ is the number of magnetic tunnel junctions (bits) per weight. We have observed that 8 bits are sufficient to store the weights with the same system precision as with real valued analog weights. $\Delta t = 10~\mathrm{\mu s}$ is the duration of the learning step. It consists on the observation phase where switches of the neurons are counted, followed by a computing phase in which equation \ref{eq:rates} is evaluated. We neglect the length of the computing phase since digital electronics is very fast compared to the observation phase. Note that if the energy barrier of the neuron junctions was lower, the switching rates would be higher and therefore the required duration of the observation phase would be lower. 

\begin{figure}
\centering
\includegraphics[width=3.5in]{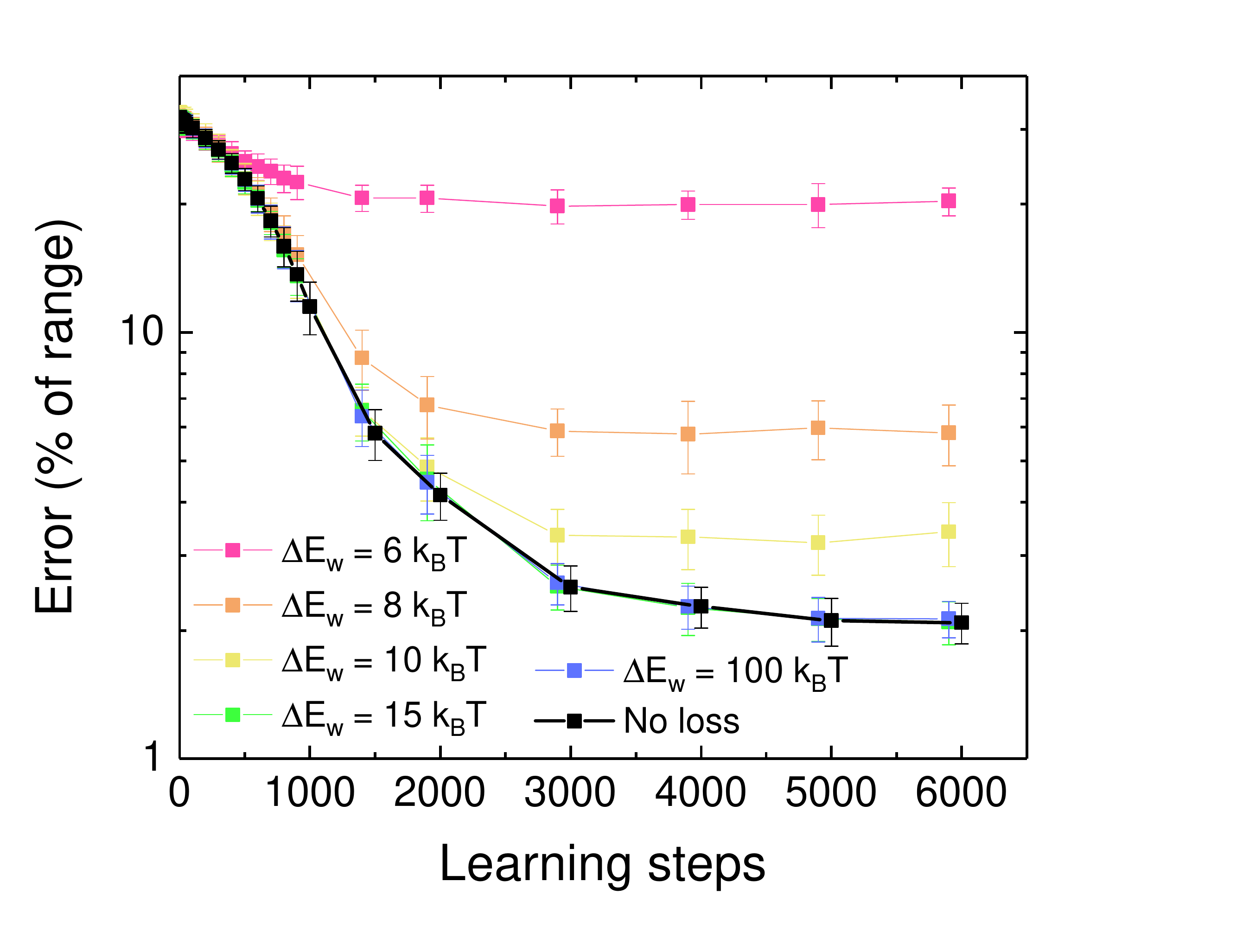}
\caption{Relative output error versus the number of learning steps, for various energy barrier heights. Each data point correspond to an average over 50 learning instances and for each point, the error was estimated as an average over 50 different input values. The error bar indicates the standard deviation, computed over the 50 learning processes. Note that the plots for $\Delta E_{\rm{w}}=15 k_{\rm{B}}T$, $\Delta E_{\rm{w}}=100 k_{\rm{B}}T$ and no loss overlap.}
\label{fig_syn_loss}
\end{figure}

Fig. \ref{fig_syn_loss} presents the evolution of the error versus the number of learning steps, for various energy barrier heights. We observe that each case reaches a steady state: after sufficient initial learning, continuous learning enables the system to adapt to dynamical changes of the weights. The steady-state error decreases with the energy barrier height (increases with the loss probability). When the energy barrier height is too low, the system is unable to learn. Energy barriers above $15 k_{\rm{B}}T$ provide as good precision as perfectly reliable weights ("no loss" label, which would correspond to an infinite energy barrier), within the statistical uncertainties. For reference, here $\Delta E_{\rm{w}}=15 k_{\rm{B}}T$ corresponds to an individual weight loss probability of $1.5 \times 10^{-7}$ per learning step. We have assumed that the unexpected reversal of one magnetic tunnel junction leads to the synaptic weight taking a random value. 

This complete randomization corresponds to a worst-case scenario as reversal of one of the less significant bits would only slightly change the weight. Furthermore, storing the weights in memristors or chained magnetic tunnel junctions would induce less drastic losses of the weights: drifts to close values rather than full randomization. Such implementations would make the system even more robust to synaptic unreliability.

\section{Power consumption versus precision tradeoff}
We compute how the unreliability of the devices storing the synaptic weights affects the power consumption of the system during continuous learning (i.e. during the steady state observed in Fig. \ref{fig_syn_loss}). The current, and therefore energy, required to write a magnetic tunnel junction scales with the critical current of the device, which is proportional to the energy barrier height.\cite{sato_properties_2014} We are thus interested in studying the trade-off between the precision of the system and the power consumption. Similar trade-offs have been considered in the context of energy efficient faulty computing.\cite{palem_ten_2013,han_approximate_2013,korkmaz_energy_2008} The power consumption of updating the synaptic weights depends on three factors: the rate at which the weights are modified during continuous learning (i.e. how often the target is missed), the number of synapses (i.e. the number of input neurons times the number of output neurons) and the power required to write a weight.

The power consumption will be proportional to the rate at which individual weights need to be updated, times the number that need updating, and the square of the energy barrier of the devices:
\begin{equation}
Power\left(N,\Delta E_\mathrm{w}\right) \propto R_{\rm update}\left(N,\Delta E_\mathrm{w}\right) \times N^2 \times \Delta E_\mathrm{w}^2
\end{equation} 
$N$ is the number of neurons in each population, therefore there are $N^2$ synapses. $R_{\rm update}$ is the rate at which the weights need to be updated.

\begin{figure}
\centering
\includegraphics[width=3.5in]{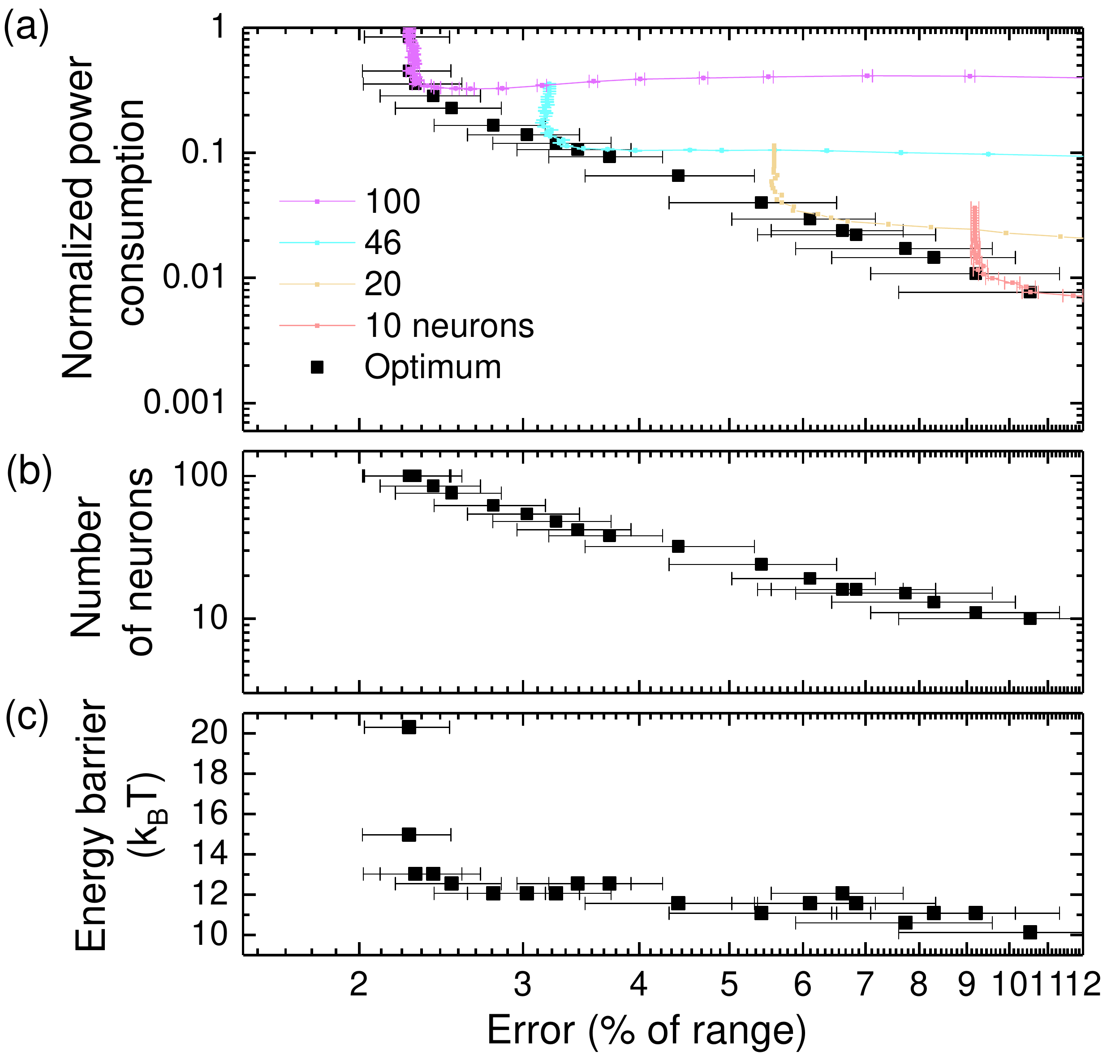}
\caption{(a) Normalized power consumption $Power\left(N,\Delta E_\mathrm{w}\right)/Power\left(100,20 k_{\rm{B}}T\right)$, plotted versus the error. Each color corresponds to the number of neurons in each population. For all curves in this figure each data point corresponds to an average over 200 learning processes and for each of them the error was estimated as an average over 50 different input values. The error bar indicates the single standard deviation of the error in the mean, computed over the 50 learning processes. For the colored curves, the error bar corresponds to a single standard deviation of the error in the mean.  Only a few of the calculations for different numbers of neurons are shown. The black squares represent the minimum power consumption versus the error. The error bars for this curve correspond to the standard deviation of the distribution.
(b)-(c) Number of neurons in each population (b) and energy barrier height (c) corresponding to the minimum power consumption to reach a given error. The error bars correspond to the standard deviation of the distribution.
}
\label{fig_tradeoff}
\end{figure}

We use numerical simulations to compute -- for various numbers of neurons and energy barrier heights -- both the error and the rate at which the weights must be updated. In each case, the power consumption is normalized by that of a system with 100 neurons in each population and an energy barrier of $\Delta E_{\rm{w}} = 20 k_{\rm{B}}T$, which in our case is effectively the same as perfectly reliable weights. Fig. \ref{fig_tradeoff} (a) shows the normalized power consumption for various barrier heights and neuron numbers (different colors). 

For each number of neurons (colored curves in Fig. \ref{fig_tradeoff}(a)) the error and the power consumption vary with  the energy barrier height. Increasing the number of neurons decreases the error because in larger systems the information carried by each synapse has a smaller relative importance. Therefore, the system is more robust to synaptic unreliability but requires greater power as there are more weights to update. Varying both the number of synapses and their reliability is necessary to minimize the power consumption required to reach a given error.  This optimal trade-off between power consumption and precision is plotted in black squares in Fig. 5(a). The minimum power consumption for a given error corresponds to a specific number of neurons and energy barrier height, which are plotted in Fig. \ref{fig_tradeoff}(b) and \ref{fig_tradeoff}(c) respectively. As expected, lower targeted errors require more neurons and more reliable synapses.

These results provide guidance on designing a population coding based systems for particular applications and neurons with particular characteristics. Given the desired precision, one can compute the optimal number of neurons as well as the optimal synaptic reliability. For example, when building a system for applications in which $3~\%$ error is acceptable, 54 stochastic magnetic tunnel junctions (neurons) in each population and magnetic tunnel junctions with an energy barrier $\Delta E = 12 k_{\mathrm{B}}T$ as weights would provide the lowest power consumption. These values of the optimal energy barriers are very low compared to the values used for traditional memory applications: retaining the information of a single magnetic tunnel junction for a few days with $99.9\%$ certainty already requires energy barriers above $40 k_{\mathrm{B}}T$.

\section{Conclusion}
Equipping a nanodevice-based population coding system with continuous learning makes it able to adapt to changes and compensate for the unreliability of its components. Continuous learning is viable as it is low power and does not require much circuit area, thus allowing on-line learning while the system is active. Our results suggest that there is a tradeoff that can be tuned between precision and power consumption. We show that for a given desired precision, it is power efficient to use unreliable weights that are energy efficient to write. When using magnetic tunnel junctions to store the weights, this translates into the possibility of using energy barriers much lower than those traditionally required for memory applications. Relaxing this constraint on the energy barrier should allow for easier device fabrication as well as improved performance with respect to other parameters.

These results are promising for hardware implementations of population coding schemes with magnetic tunnel junctions as neurons and synaptic weights. They should remain valid qualitatively independent of the devices used, highlighting the idea that using brain inspired approaches, such as continuous learning can allow a system with unreliable nanodevices to perform reliably.

\begin{acknowledgments}
A.M. acknowledges support under the Cooperative Research Agreement between the University of Maryland and the National Institute of Standards and Technology, Center for Nanoscale Science and Technology, Grant No. 70NANB10H193, through the University of Maryland.
This work was supported in part by the European Research Council ERC under Grant bioSPINspired 682955.
\end{acknowledgments}

\section*{Appendix: Details of the simulations}

For all simulations, the evolution of the state of each neuron junction (in both the input and output populations) is simulated using Eq. \ref{eq:proba} and \ref{eq:model_smtj}. The energy barrier of the neuron junctions $\Delta E_{\rm{n}}$ is so that $\frac{\Delta E_{\rm{n}}}{k_{\rm B}T}=6$ and the critical voltage is $V_{\rm{c}}=0.1\mathrm{V}$. The input voltage $V$ can take values between $-0.1~\mathrm{V}$ and $0.1~\mathrm{V}$. 
In each neuron population, the junction of index $i$ receives an individual bias voltage $V_{0i}$ spanning from $-0.15~\mathrm{V}$ to $0.15~\mathrm{V}$ across the population. The rates of the neurons in each populations are computed by counting the number of switches over $10~\mu \mathrm{s}$. Computing the rate by counting stochastic switches makes it intrinsically inexact. The observation length leads to a tradeoff between energy consumption and precision as a longer observation consumes more but provides a better estimation of the rate. This tradeoff is typical of stochastic computing.\cite{alaghi_survey_2013} Here, for simplicity, we keep the observation length fixed through the paper. 

The voltage $V_\mathrm{out}$ corresponding to the output of the system is deduced from them by a standard population decoding technique:\cite{salinas_transfer_1995}
\begin{equation}
V_\mathrm{out}=\frac{\sum_{i=1}^{N_\mathrm{out}}V_{0i}R^\mathrm{out}_i}{\sum_{i=1}^{N_\mathrm{out}}R^\mathrm{out}_i}
\end{equation}

The expression of the desired target output is:
\begin{equation}
V_{\rm{Target}} = V_{\rm{max}} \times sin(V_{\rm{Input}}/(V_{\rm{max}}  \pi) )
\end{equation}
where $V_{\rm{max}}=0.1~\mathrm{V}$ being the boundary of the input voltage range. \\

At each learning step, the learning rule is implemented as follows. If the output matches the target within a chosen range (i.e. the size of the target), the weights are not modified. In this paper we chose the size of the target to be $5\%$ of the possible output values range. If the output is higher than the target: the weights connecting the input population to the junctions which bias voltages correspond to values higher than the output are decreased and the weights connecting the input population to the junctions which bias voltages correspond to values lower than the output are increased. If, in contrast, the output is lower than the target, the opposite is implemented.

The increase or decrease of the weights connecting the $j$-th output junction to the input population corresponds to:
\begin{equation}
W_{ij} \rightarrow \left( W_{ij} \pm \alpha \frac{R^\mathrm{in}_i}{f_0} \right) \frac{1}{1+\alpha}
\end{equation}
where $\alpha=0.001$ is the learning rate.

\bibliographystyle{apsrev4-1}
\bibliography{Zotero}

\end{document}